%% file: main.tex
\documentclass[runningheads]{llncs}

\newif\ifextended
\extendedtrue 

\input{packages}
\input{commands}

\begin{document}
\title{BDDs Strike Back
\thanks{This work has been partially funded by NWO under the grant \href{https://primavera-project.com}{PrimaVera} number NWA.1160.18.238,
 European Union’s Horizon 2020 research and innovation programme under the Marie Skłodowska-Curie grant agreement No 101008233 (\emph{Mission}), and
the ERC Consolidator Grant 864075 (\emph{CAESAR}).
Khan is funded by a HEC-DAAD stipend.}}
\subtitle{Efficient Analysis of Static and Dynamic Fault Trees}
%
%
\author{Daniel~Basg{\"o}ze\inst{1} \and
Matthias~Volk\inst{2}\orcidID{0000-0002-3810-4185} \and
Joost-Pieter~Katoen\inst{1,2}\orcidID{0000-0002-6143-1926} \and
Shahid~Khan\inst{1}\orcidID{0000-0001-5549-7809} \and
Marielle~Stoelinga\inst{2,3}\orcidID{0000-0001-6793-8165}}
\authorrunning{D. Basg{\"o}ze et al.}
\titlerunning{Faster fault tree analysis using BDDs}
%
\institute{%
Software Modeling and Verification, RWTH Aachen University, Aachen, Germany \and
Formal Methods and Tools, University of Twente, Enschede, The Netherlands
\email{m.volk@utwente.nl} \and
Department of Software Science, Radboud University, Nijmegen, The Netherlands}
\maketitle

\input{sections/abstract}
\input{sections/introduction}
\input{sections/preliminaries}
\input{sections/sft-bdd}

\input{sections/evaluation_sft}
\input{sections/dft-bdd}
\input{sections/evaluation_dft}

\input{sections/conclusion}

\bibliographystyle{splncs04}
\bibliography{bibliography}
%

\ifextended
    \clearpage
    \appendix
    \input{sections/appendix}
\fi
\end{document}

%% file: commands.tex
\newcommand{\ie}{i.e., }
\newcommand{\eg}{e.g., }
\newcommand{\wrt}{w.r.t.\ }
\newcommand{\cf}{cf.\ }

\newcommand{\tool}[1]{\textsc{#1}\xspace}
\newcommand{\storm}{\tool{Storm}}
\newcommand{\stormdft}{\tool{Storm-dft}}
\newcommand{\sylvan}{\tool{Sylvan}}
\newcommand{\scram}{\tool{Scram}}
\newcommand{\xfta}{\tool{Xfta}}
\newcommand{\ffort}{\tool{FFORT}}

\newcommand*{\dftType}[1]{\textsf{#1}\xspace}
\newcommand*{\BE}{\dftType{BE}}

\newcommand*{\AND}{\dftType{AND}}
\newcommand*{\OR}{\dftType{OR}}
\newcommand*{\VOT}{\dftType{VOT}}
\newcommand*{\VOTk}[1]{\ensuremath{\VOT_{#1}}\xspace}

\newcommand*{\PAND}{\dftType{PAND}}

\newcommand*{\POR}{\dftType{POR}}

\newcommand*{\SPARE}{\dftType{SPARE}}
\newcommand*{\FDEP}{\dftType{FDEP}}
\newcommand*{\PDEP}{\dftType{PDEP}}
\newcommand*{\PDEPp}[1]{\ensuremath{\PDEP_{#1}}\xspace}
\newcommand*{\SEQ}{\dftType{SEQ}}

\newcommand*{\dft}{\ensuremath{\mathcal{F}}\xspace}
\newcommand*{\nodes}{\ensuremath{V}}
\newcommand*{\childrenS}{\sigma}

\newcommand*{\typeS}{\textit{Type}}
\newcommand*{\type}[1]{\typeS(#1)}
\newcommand*{\tle}{\textit{top}\xspace}
\newcommand*{\attachSE}{\Theta}
\newcommand*{\dftTuple}{\tuple{V,\childrenS,\typeS,\tle,\attachSE}}

\newcommand*{\set}[1]{\left\{#1\right\}}
\newcommand*{\tuple}[1]{\left(#1\right)}

\newcommand*{\Distributions}{\Omega}
\newcommand*{\prob}[2][]{\textsf{P}_{#1}(#2)}

\newcommand{\ctmc}{\ensuremath{\mathcal{C}}}

\newcommand*{\Var}{\text{Var}}

\newcommand*{\cof}[2]{{#1|}_{#2}}
\newcommand*{\bdd}{\ensuremath{\mathcal{B}}}

\newcommand*{\varord}{\ensuremath{<_{\Var}}}

\newcommand{\BI}{\textrm{BI}}

%% file: sections/abstract.tex
\begin{abstract}
Fault trees are a key model in reliability analysis.
Classical static fault trees (SFT) can best be analysed using binary decision diagrams (BDD).
State-based techniques are favorable for the more expressive dynamic fault trees (DFT).
This paper combines the best of both worlds by following Dugan's approach: dynamic sub-trees are analysed via model checking Markov models and replaced by basic events capturing the obtained failure probabilities.
The resulting SFT is then analysed via BDDs.
We implemented this approach in the \storm model checker.
Extensive experiments (a) compare our pure BDD-based analysis of SFTs to various existing SFT analysis tools, (b) indicate the benefits of our efficient calculations for multiple time points and the assessment of the mean-time-to-failure, and (c) show that our implementation of Dugan's approach significantly outperforms pure Markovian analysis of DFTs.
Our implementation \stormdft is currently the only tool supporting efficient analysis for both SFTs and DFTs.
\end{abstract}

%% file: sections/introduction.tex
\section{Introduction}
\label{sec:introduction}

\emph{Fault trees}~\cite{FaultTreeHandbook02,DBLP:journals/csr/RuijtersS15,DBLP:books/cu/TrivediB17} are a common formalism in reliability engineering and required by standards in a broad range of industries~\cite{FaultTreeHandbook02,FAA00,iso26262}.
A fault tree represents a Boolean function and models how overall system failures depend on the failure of basic system components.
Fault tree analysis (FTA) is commonly performed by translating a fault tree into a binary decision diagram (BDD) and calculating the relevant metrics on this BDD~\cite{rauzy1993new,sinnamon1996fault}.
BDDs yield compact representations of fault trees enabling the analysis of large systems~\cite{coudert1993fault}.
Ongoing improvements in BDD tools such as parallelisation~\cite{DBLP:journals/sttt/DijkP17} allow for modern implementations of FTA via BDDs.

Fault trees are static in nature and their expressiveness is limited.
\emph{Dynamic fault trees (DFT)}~\cite{DuganBB90} support ordered failures, spare management and functional dependencies.
The flexibility and increased expressiveness of DFTs however requires more involved analysis methods.
BDDs cannot represent DFTs directly as DFTs consider failure sequences instead of Boolean combinations.
A common approach is to translate DFTs into Markov models~\cite{dugan1992dynamic,DBLP:journals/tdsc/BoudaliCS10,DBLP:journals/tii/VolkJK18}.
Gulati and Dugan~\cite{gulati1997modular} proposed a modular DFT analysis approach combining several analysis techniques.
It divides the DFT into independent sub-parts which are analysed individually.
\emph{Modularisation thus allows to use the best of both worlds: Markov models for dynamic parts and BDDs for static parts.}

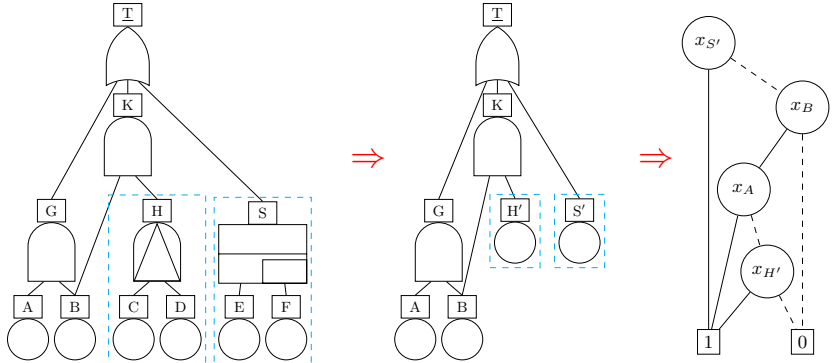
\begin{figure}[t]
    \centering
    \scalebox{0.78}{\input{figures/ex_dft_modular.tex}}
    \caption{DFT modularisation}
    \label{fig:ex_dft_modular}
\end{figure}
The idea of modularisation is depicted in Fig.~\ref{fig:ex_dft_modular}.
First, dynamic modules, \ie sub-trees containing dynamic elements, are identified (the blue boxes in the left-most tree).
Each dynamic module is analysed independently with state-of-the-art analysis techniques for Markov models~\cite{DBLP:journals/tii/VolkJK18}.
Afterwards, each dynamic module is replaced by a single basic event which represents the corresponding failure probabilities (tree in the middle).
The remaining (static) fault tree is then translated into a BDD (right most part) and is analysed by BDD techniques.
Modularisation works especially well when the dynamic parts are contained at the bottom of the fault tree and the static parts are on top.
As dynamic parts commonly model single components, this structure is present in most DFTs.

\paragraph{Related work.}
FTA via BDDs was first presented in~\cite{rauzy1993new} and~\cite{coudert1993fault}, and successively improved in~\cite{sinnamon1996fault,DBLP:journals/ress/DutuitR01}.
BDD-based analysis of static fault trees (SFTs) is supported by academic tools such as \scram~\cite{scram18}, \xfta~\cite{rauzy2020probabilistic} and \tool{SHARPE}~\cite{DBLP:journals/sigmetrics/TrivediS09}, as well as commercial tools, \eg \tool{RiskSpectrum}~\cite{backstrom2018experimental}.
We refer to~\cite{DBLP:journals/csr/RuijtersS15} for a detailed overview on BDD-based FTA.

For DFTs, various analysis techniques exist.
Common approaches translate DFTs into models such as Markov models~\cite{dugan1992dynamic,DBLP:journals/tdsc/BoudaliCS10,DBLP:journals/tii/VolkJK18}, Bayesian networks~\cite{DBLP:journals/ress/MontaniPBC08} and Petri nets~\cite{DBLP:journals/entcs/Raiteri05}, or are based on Monte-Carlo simulation~\cite{DBLP:journals/ress/RaoGRKVS09,DBLP:conf/qest/BuddeRS20}.
There also exist analysis techniques based on extensions of BDDs such as sequence decision diagrams~\cite{DBLP:journals/ress/Rauzy11}, sequential BDDs~\cite{DBLP:journals/tsmc/XingTD12}, multiple-valued decision diagrams~\cite{DBLP:journals/tr/Mo14} or conditional BDDs~\cite{DBLP:journals/jss/ZhouXW20}.
These approaches use BDDs to enumerate all failure sequences leading to a system failure so as to compute the overall system unreliability.
To the best of our knowledge, no tool support exists for any of the BDD approaches for DFTs and their scalability for large DFTs remains unclear.

For SD fault trees~\cite{DBLP:conf/dsn/KrcalK15}---another extension of static fault trees with limited dynamic behaviour---efficient analysis techniques exist via minimal cut sets~\cite{DBLP:conf/dsn/KrcalK15} and abstraction~\cite{DBLP:conf/safecomp/BackstromBHKK16}.
However, expressiveness of SD fault trees is limited compared to DFTs as they only allow dynamic behaviour in basic system components.

Our DFT approach is based on modularisation~\cite{gulati1997modular} which was first implemented in the \tool{DIFtree} tool~\cite{dugan1997diftree} and its successor \tool{Galileo}~\cite{DBLP:conf/ftcs/SullivanDC99}.
However, both tools are not available anymore for more than a decade.
Recent work~\cite{DBLP:journals/tii/VolkJK18} has implemented modularisation for DFTs but is limited to independent sub-trees that must be direct children of the top event.
In addition, \cite{DBLP:journals/tii/VolkJK18} analyses static FTs by translation to Markov models.

\paragraph{Implementation.}
We implemented the BDD translation for static fault trees (SFTs) in the \storm model checker~\cite{hensel2021probabilistic} and use the multi-core BDD library \sylvan~\cite{DBLP:journals/sttt/DijkP17}.
Our implementation \stormdft supports computing minimal cut sets (MCS), the unreliability and several importance measures such as the Birnbaum index~\cite{birnbaum1969importance}.
\stormdft exploits vectorisation and thus enables to compute a metric for multiple time bounds at once.
In addition, we support the calculation of the mean-time-to-failure (MTTF) via approximation.
For DFTs, we implemented the modularisation approach exploiting both the BDD translation and our efficient DFT analysis via Markov models~\cite{DBLP:journals/tii/VolkJK18}.

\paragraph{Evaluation.}
Experiments on a benchmark set of $215$ SFTs and $124$ DFTs yield:
\begin{compactitem}
    \item On SFTs, \stormdft is competitive compared to existing tools such as \scram~\cite{scram18} and \xfta~\cite{rauzy2020probabilistic} and is significantly faster when analysing multiple time points due to vectorisation.
    \item The variable ordering in \stormdft is not yet optimal and can lead to larger BDDs and subsequently longer run times.
    \item On DFTs, our BDD-based modularisation is significantly faster than both plain Markov-model analysis~\cite{DBLP:journals/tii/VolkJK18} and an existing realisation of (top-down) modularisation~\cite{DBLP:journals/tii/VolkJK18} based on pure Markov-model analysis.
\end{compactitem}

Our implementation is publicly available in the open-source tool \stormdft\footnote{\url{https://www.stormchecker.org/}}.
We also provide an artifact of the experimental evaluation containing the scripts, tool configurations and fault tree models\footnote{\url{https://doi.org/10.5281/zenodo.6390998}}.

\paragraph{Contributions.}
In summary, the main contributions of this paper are:
\begin{compactitem}
    \item A competitive implementation of SFT analysis via BDDs in \storm~\cite{hensel2021probabilistic}.
    \item Vectorisation for multiple time bounds and an approximation for MTTF.
    \item A fast implementation of a modern version of modularisation using BDDs.
\end{compactitem}
\emph{Our implementation \stormdft is the only state-of-the-art analysis tool for both static and dynamic fault trees.}

\paragraph{Structure of the paper.}
We introduce fault trees and binary decision diagrams in Sect.~\ref{sec:prelim}.
Section~\ref{sec:sft-bdd} presents the analysis of (static) fault trees using BDDs.
We evaluate the approach in Sect.~\ref{sec:evaluation_sft}.
Section~\ref{sec:dft-bdd} presents the analysis of dynamic fault trees via modularisation and using BDDs for static parts.
We evaluate the approach in Sect.~\ref{sec:evaluation_dft}.
We conclude in Sect.~\ref{sec:conclusion} and present future work.

%% file: figures/ex_dft_modular.tex
\begin{tikzpicture}[bdd]
	\node[or3] (T) at (6.5, 1.3) {};
	\node[labelbox] (T_label) at (T.east) {\underline{T}};
	\node[and2] (K) at (6.5, -0.2) {};
	\node[labelbox] (K_label) at (K.east) {K};
	\node[and2] (G) at (5.2, -2.0) {};
	\node[labelbox] (G_label) at (G.east) {G};
	\node[and2] (H) at (7.0, -2.0) {};
	\node[triangle_pand] (triangle_H) at(H) {};
	\node[labelbox] (H_label) at (H.east) {H};
	\node[spare] (S) at (8.8, -2.0) {};
	\node[labelbox] (S_label) at (S.north) {S};
	\node[be] (A) at (4.8, -3.1) {};
	\node[labelbox] (A_label) at (A.north) {A};
	\node[be] (B) at (5.6, -3.1) {};
	\node[labelbox] (B_label) at (B.north) {B};
	\node[be] (C) at (6.6, -3.1) {};
	\node[labelbox] (C_label) at (C.north) {C};
	\node[be] (D) at (7.4, -3.1) {};
	\node[labelbox] (D_label) at (D.north) {D};
	\node[be] (E) at (8.4, -3.1) {};
	\node[labelbox] (E_label) at (E.north) {E};
	\node[be] (F) at (9.2, -3.1) {};
	\node[labelbox] (F_label) at (F.north) {F};

	\draw[-](T.input 1) -- (G_label.north);
	\draw[-](T.input 2) -- (K_label.north);
	\draw[-](T.input 3) -- (S_label.north);
	\draw[-](K.input 1) -- (B_label.north);
	\draw[-](K.input 2) -- (H_label.north);
	\draw[-](G.input 1) -- (A_label.north);
	\draw[-](G.input 2) -- (B_label.north);
	\draw[-](H.input 1) -- (C_label.north);
	\draw[-](H.input 2) -- (D_label.north);
	\draw[-](S.P) -- (E_label.north);
	\draw[-](S.SC) -- (F_label.north);

    \node[draw=cyan, dashed, inner sep=2pt, fit=(H_label) (C) (D)] {};
    \node[draw=cyan, dashed, inner sep=2pt, fit=(S_label) (E) (F)] {};

    \node[] (arrow1) at (10.6, -0.4) {\LARGE\textcolor{red}{$\Rightarrow$}};

	\node[or3] (T) at (12.8, 1.3) {};
	\node[labelbox] (T_label) at (T.east) {\underline{T}};
	\node[and2] (K) at (12.8, -0.2) {};
	\node[labelbox] (K_label) at (K.east) {K};
	\node[and2] (G) at (11.8, -2.0) {};
	\node[labelbox] (G_label) at (G.east) {G};
	\node[be] (H) at (13.1, -1.45) {};
	\node[labelbox] (H_label) at (H.north) {H$^\prime$};
	\node[be] (S) at (14.2, -1.45) {};
	\node[labelbox] (S_label) at (S.north) {S$^\prime$};
	\node[be] (A) at (11.4, -3.1) {};
	\node[labelbox] (A_label) at (A.north) {A};
	\node[be] (B) at (12.2, -3.1) {};
	\node[labelbox] (B_label) at (B.north) {B};

	\draw[-](T.input 1) -- (G_label.north);
	\draw[-](T.input 2) -- (K_label.north);
	\draw[-](T.input 3) -- (S_label.north);
	\draw[-](K.input 1) -- (B_label.north);
	\draw[-](K.input 2) -- (H_label.north);
	\draw[-](G.input 1) -- (A_label.north);
	\draw[-](G.input 2) -- (B_label.north);

    \node[draw=cyan, dashed, inner sep=2pt, fit=(H_label) (H)] {};
    \node[draw=cyan, dashed, inner sep=2pt, fit=(S_label) (S)] {};

    \node[] (arrow2) at (15.5, -0.4) {\LARGE\textcolor{red}{$\Rightarrow$}};

    \node[innernode, minimum size=0.9cm] at (16.4,1.6) (s) {$x_{S^\prime}$};
    \node[innernode, minimum size=0.9cm] at (18.0,0.5) (b) {$x_B$};
    \node[innernode, minimum size=0.9cm] at (17.0,-0.9) (a) {$x_{A}$};
    \node[innernode, minimum size=0.9cm] at (17.4,-2.3) (h) {$x_{H^\prime}$};
	\node[leaf] at (16.4,-3.5) (one) {$1$};
	\node[leaf] at (18.0,-3.5) (zero) {$0$};

	\draw[succone]
	  (s) edge (one) {}
      (b) edge (a) {}
	  (a) edge (one) {}
	  (h) edge (one) {}
	;
    \path[succzero]
	  (s) edge (b) {}
	  (b) edge (zero) {}
      (a) edge (h) {}
	  (h) edge (zero) {}
	;
\end{tikzpicture}

%% file: sections/preliminaries.tex
\section{Preliminaries}
\label{sec:prelim}

\subsection{Fault trees}
\label{subsec:fts}
Fault trees (FTs) model how failures can occur and propagate in systems~\cite{FaultTreeHandbook02,DBLP:journals/csr/RuijtersS15,DBLP:books/cu/TrivediB17}.
FTs are \emph{directed acyclic graphs} in which the leaves are called \emph{basic events} (\BE{}s) and intermediate nodes are called \emph{gates}.
A \BE{} represents an (atomic) system component which fails according to a given failure distribution.
Failures of BEs are propagated through the system according to the gates and eventually lead to a failure of the unique root of the graph, the \emph{top event}.
\emph{Dynamic fault trees (DFTs)}~\cite{DuganBB90} are the most prominent extension of fault trees and their additional gates allow for more realistic modelling.
Note that we do not consider repairs.

\begin{definition}[Dynamic fault tree]
    A \emph{dynamic fault tree (DFT)} is a tuple $\dft = \dftTuple$ where
	\begin{compactitem}
		\item $\nodes$ is a finite set of \emph{nodes}.
        \item $\childrenS: \nodes \rightarrow \nodes^{*}$ defines the \emph{ordered children} of a node (also called the inputs).
        \item $\typeS: \nodes \rightarrow \set{\BE, \AND, \OR, \VOTk{k}} \cup \set{\PAND, \POR, \PDEP, \SEQ, \SPARE}$ defines the \emph{type of a node}.
		\item $\tle \in \nodes$ is the \emph{top event}.
        \item $\attachSE: \set{v \in \nodes \mid \type{v} = \BE} \to \Distributions$ maps each \BE to a \emph{failure distribution} from $\Distributions$ the set of probability distributions.
	\end{compactitem}
\end{definition}
A $\VOTk{k}$-gate satisfies $1\leq k \leq |\childrenS|$.
A \emph{static fault tree (SFT)} is a DFT where the node types $\typeS$ are restricted to $\set{\BE, \AND, \OR, \VOTk{k}}$.
In DFTs, we restrict the failure distributions of \BE{}s to exponential distributions to allow for analysis based on Markov models.
We say a ``DFT $\dft$ is failed'' if $\tle$ is failed.

\input{figures/dft_elements}
We shortly introduce the different node types; the precise semantics is given in~\cite{DBLP:conf/apn/JungesKS018}.
The graphical representation of each node type is given in Fig.~\ref{fig:DFTElements}.

\emph{Basic events} (\BE{}s) fail according to their associated failure distributions.
Commonly, an exponential failure distribution with a failure rate $\lambda$ is used.

\emph{Static gates} represent Boolean logic functions.
The \AND-gate fails if all its inputs fail.
The \OR-gate fails if at least one input fails.
The \VOTk{k}-gate is a generalisation and fails if at least $k$ inputs fail.

\emph{Priority gates} extend the static gates with the additional constraint that the inputs have to fail in order from left to right.
Failures out of this order render the gate \emph{fail-safe} and it can never fail.
The \PAND-gate fails if all inputs fail from left to right.
The \POR-gate fails if the leftmost child fails before all other gates.

\emph{Dependencies} encode functional dependencies of the system.
If the first child of the \PDEPp{p} fails, all other children fail with probability $p$.

\emph{Sequence enforcers} ensure that children only fail in order from left to right.

\emph{Spare gates} model spare management.
Initially, the first child is used.
If it fails, the next child is claimed and used, and so forth.
The \SPARE fails if all its children failed.
Children can be shared by multiple \SPARE{}s, but can only be used exclusively by one \SPARE.
Using a child activates the corresponding components and can increase the associated failure rate.

\begin{example}[Fault tree]
    Consider the DFT on the left of Fig.~\ref{fig:ex_dft_modular}.
    The top event~$T$ fails for example if both \BE{}s $A$ and $B$ fail.
    The DFT also fails if $B$ and \PAND $H$ fail.    
    The \PAND only fails if the first child $C$ fails before $D$.
\end{example}

\paragraph{Fault tree analysis.}
Fault trees are analysed \wrt the failure of the top event.
Common metrics are the \emph{unreliability} within a given time bound and the \emph{mean-time-to-failure (MTTF)}.
For SFTs, the \emph{minimal cut sets} play an important role.
Cut sets with only a few elements for example indicate system vulnerabilities.

\begin{definition}[Minimal cut sets]
    \label{def:mcs}
    Let $\dft$ be a \emph{static} fault tree.
    A \emph{minimal cut set (MCS)} for $\dft$ is a set $M \subseteq \BE{}$ such that:
    \begin{compactenum}
    \item the failure of all \BE{}s in $M$ leads to the failure of $\dft$, and
        \item $M$ is minimal, \ie no subset $M' \subsetneq M$ leads to the failure of $\dft$.
    \end{compactenum}
\end{definition}

\subsection{Binary decision diagrams}
\label{subsec:bdds}
\emph{Binary decision diagrams (BDDs)}~\cite{DBLP:journals/tc/Akers78,DBLP:journals/tc/Bryant86} are graphs based on the Shannon expansion~\cite{shannon1938}.
We introduce BDDs by example and refer to~\cite{DBLP:reference/mc/Bryant18} for more details.

A BDD $\bdd$ encodes a Boolean function $f$ over variables $x_1, \dots, x_n$.
Nodes in $\bdd$ represent variables of $f$ and follow a given variable ordering.
Outgoing edges of a node $x$ represent the two possible assignments of variable $x$: the solid line represents $x=1$, the dashed line $x=0$.
Leaves represent functions $1$ and $0$.

\begin{example}[BDD]
    Consider the BDD on the right of Fig.~\ref{fig:ex_dft_modular}.
    The BDD represents the function $f= x_{S'} \vee \left(x_{B} \wedge \left(x_{A} \vee x_{H'}\right)\right)$.
    The satisfying assignments of $f$ are obtained by following all paths from the root to the $1$-leaf.
\end{example}

%% file: figures/dft_elements.tex
\begin{figure}[t]
\centering
\subfigure[\BE]{
 \centering
\makebox[0.060\linewidth]{
\scalebox{0.62}{
 \begin{tikzpicture}[  scale=.8,font=\LARGE,text=black, every node/.style={transform shape}, node distance=0.3cm]
	\node[be] (pand) {};
	\node[above=of pand] (output) {};
	\draw[-] (pand) -- (output);
\end{tikzpicture}}}
 \label{fig:BE}
}
\subfigure[$\AND$]{
  \centering
\makebox[0.083\linewidth]{
\scalebox{0.62}{
 \begin{tikzpicture}
    \node[and3] (and) {};
    \node[below=0.4 cm of and.input 1, xshift=-0.2cm]  (i1) {};
    \node[below=0.3 cm of and.input 2]  (dots) {$\hdots$};
    
    \node[below=0.4 cm of and.input 3, xshift=0.2cm] (i2) {};
    
    \draw[-] (and.input 1) -- (i1);
    \draw[-] (and.input 3) -- (i2);
  \end{tikzpicture}
  }
  }
 \label{fig:AND} 
}
\subfigure[$\OR$]{
  \centering
\makebox[0.065\linewidth]{
\scalebox{0.62}{
 \begin{tikzpicture}
    \node[or3] (and) {};
    \node[below=0.4 cm of and.input 1, xshift=-0.2cm]  (i1) {};
    \node[below=0.3 cm of and.input 2]  (dots) {$\hdots$};
    
    \node[below=0.4 cm of and.input 3, xshift=0.2cm] (i2) {};
    
    \draw[-] (and.input 1) -- (i1);
    \draw[-] (and.input 3) -- (i2);
  \end{tikzpicture}
  }
  }
 \label{fig:OR}
}
\subfigure[$\VOTk{k}$]{
 \centering
\makebox[0.095\linewidth]{
\scalebox{0.62}{
 \begin{tikzpicture}
    \node[and3] (and) {\rotatebox{270}{$k$}};
    \node[below=0.4 cm of and.input 1, xshift=-0.2cm]  (i1) {};
    \node[below=0.3 cm of and.input 2]  (dots) {$\hdots$};
    
    \node[below=0.4 cm of and.input 3, xshift=0.2cm] (i2) {};
    
    \draw[-] (and.input 1) -- (i1);
    \draw[-] (and.input 3) -- (i2);
  \end{tikzpicture}}}
 \label{fig:VOT}
}
 \subfigure[\PAND]{
 \centering
\makebox[0.095\linewidth]{
 \scalebox{0.62}{
   \begin{tikzpicture}   
    \node[and3] (and) {{\tiny\rotatebox{270}{$\leq$}}};
    \node[triangle,scale=1.62,yshift=-3.5,xscale=0.80] (triangle_a) at (and) {};
    \node[below=0.4 cm of and.input 1, xshift=-0.2cm]  (i1) {};
    \node[below=0.3 cm of and.input 2]  (dots) {$\hdots$};
    
    \node[below=0.4 cm of and.input 3, xshift=0.2cm] (i2) {};
    
    \draw[-] (and.input 1) -- (i1);
    \draw[-] (and.input 3) -- (i2);
  \end{tikzpicture}}}
  \label{fig:DFTElements_PAND}
  
 }
\subfigure[\POR]{
  \centering
\makebox[0.079\linewidth]{
  \scalebox{0.62}{
    \begin{tikzpicture}   
    \node[or2] (and) {{\tiny\rotatebox{270}{$\leq$}}};
    \node[btriangle,scale=1.61,yscale=0.915, xshift=-0.113cm] (triangle_b) at (and) {};
    \node[below=0.4 cm of and.input 1]  (i1) {};
    
    \node[below=0.4 cm of and.input 2] (i2) {};
    
    \draw[-] (and.input 1) -- (i1);
    \draw[-] (and.input 2) -- (i2);
  \end{tikzpicture}}}
  \label{fig:DFTElements_POR}
 }
 \subfigure[\PDEP]{
   \centering
   \scalebox{0.62}{
      \begin{tikzpicture}   
    
    \node[fdep] (and) {};
    \node[above=0.07cm of and.center] (x) {$p$};
    \node[below=0.7 cm of and.T, xshift=-0.2cm]  (i1) {};
    
    \node[below=0.4 cm of and.EB] (i2) {};
    \draw[-] (and.T) -- (i1);
    \draw[-] (and.EB) -- (i2);
    
  \end{tikzpicture}
  }
    \label{fig:DFTElements_PDEP}
 }
 \subfigure[\SEQ]{
  \centering
  \makebox[0.082\linewidth]{
  \scalebox{0.62}{
    \begin{tikzpicture}   
    \node[seq] (and) {$\rightarrow$};
    \node[below=0.4 cm of and.250, xshift=-0.2cm]  (i1) {};
    \node[below=0.3 cm of and.270]  (dots) {$\hdots$};
    
    \node[below=0.4 cm of and.290, xshift=0.2cm] (i2) {};
    
    \draw[-] (and.250) -- (i1);
    \draw[-] (and.290) -- (i2);
  \end{tikzpicture}}}
  \label{fig:DFTElements_SEQ}
 }
 \subfigure[\SPARE]{
\centering
\makebox[0.110\linewidth]{
\scalebox{0.62}{
  \begin{tikzpicture}   
    \node[spare] (and) {};
    \node[below=0.4 cm of and.P]  (i1) {};

    \node[below=0.4 cm of and.SA] (i2) {};
    \node[below=0.3 cm of and.SC] (i3) {$\hdots$};
    
    \node[below=0.4 cm of and.SE, xshift=0.3cm] (i4) {};
    
    \draw[-] (and.P) -- (i1);
    \draw[-] (and.SA) -- (i2);
    \draw[-] (and.SE) -- (i4);
  \end{tikzpicture} }}
  \label{fig:DFTElements_SPARE}
 }
 \caption{Node types in ((a)-(d)) static and (all) dynamic fault trees.} 
 \label{fig:DFTElements}
\end{figure}
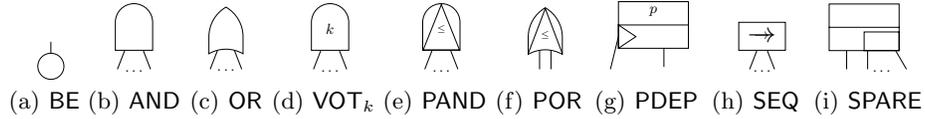%

%% file: sections/sft-bdd.tex
\section{SFT analysis via BDD}
\label{sec:sft-bdd}

\paragraph{Translation from SFT into BDD.}
An SFT $\dft$ is translated into a BDD by simply calculating the BDD-representation of the propositional formula representing the failure behaviour of $\dft$~\cite{rauzy1993new}.
The algorithm follows a simple recursive bottom-up approach which combines the BDDs of sub-trees according to the logic gates.
The \VOTk{k}-gate is translated by exploiting the Shannon decomposition~\cite{shannon1938}.

\paragraph{Variable ordering.}
The variable ordering employed for the BDD is important as different orderings can result in significantly different BDD sizes~\cite{sinnamon1997improved,andrews1998efficient}.
Common variable orderings are \emph{depth-first search (DFS)}~\cite{rauzy1993new} and \emph{top-down left-right (TDLR)}~\cite{andrews1998efficient}.
Finding optimal variable orderings is still ongoing research~\cite{bouissou1997bdd,rauzy2008some}.

\begin{figure}[t]
    \centering
    \subfigure[Sample SFT]{
        \centering
        \scalebox{0.9}{\input{figures/ex_sft.tex}}
        \label{subfig:sft}
    }
    \hfill
    \subfigure[BDD with DFS]{
        \centering
        \scalebox{0.9}{\input{figures/ex_bdd_dfs.tex}}
        \label{subfig:bdd_dfs}
    }
    \hfill
    \subfigure[BDD with TDLR]{
        \centering
        \makebox[0.3\linewidth]{
        \scalebox{0.9}{\input{figures/ex_bdd_tdlr.tex}}}
        \label{subfig:bdd_tdlr}
    }
    \caption{SFT and corresponding BDDs for different variable orders}
    \label{fig:sft_bdd}
\end{figure}
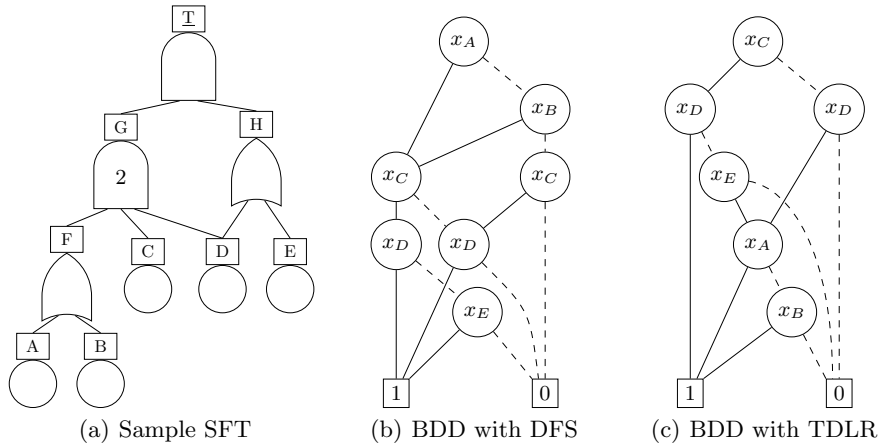

\begin{example}[BDDs for SFT]
    Consider the SFT depicted in Fig.~\ref{subfig:sft}.
    The corresponding BDD for the DFS ordering $x_A \varord x_B \varord x_C \varord x_D \varord x_E$ has 7 nodes and is given in Fig.~\ref{subfig:bdd_dfs}.
    The BDD for variable ordering TDLR $x_C \varord x_D \varord x_E \varord x_A \varord x_B$ has 6 nodes and is depicted in Fig.~\ref{subfig:bdd_tdlr}.
\end{example}

\subsection{Computing minimal cut sets}
Minimal cut sets are a common metric for SFTs.
Several approaches exist to compute MCS~\cite{DBLP:journals/csr/RuijtersS15}, and we focus on the BDD-based approach~\cite{rauzy1993new}.
We are interested in all paths of the BDD that reach the $1$-leaf, \ie lead to a failure of the SFT.
All variables reached by the $1$-edge on such a path form a \emph{solution} of the BDD.
Each solution is a cut set.
The aim is to compute all minimal solutions, \ie solutions whose proper subsets are not a solution.

A na\"ive approach computes the solutions for each (sub-)BDD in a bottom-up way.
The solutions of a node $v$ are then the union of the solutions of the $1$-successor of $v$ extended with $v$, and the solutions of the $0$-successor.
However, the resulting solutions are not necessarily minimal~\cite{rauzy1993new,Basgoeze20}.

The algorithm of~\cite{rauzy1993new} exploits the fact that SFTs can only encode monotonic switching functions (a.k.a.\ \emph{coherent} FTs).
From the original BDD $\bdd$, a new BDD $\bdd'$ is constructed whose solutions are exactly the minimal solutions of $\bdd$.
This construction uses the `without'-operator on BDDs to exclude parts of the $1$-successor which are already included in the $0$-successor;
see~\cite{rauzy1993new,Basgoeze20} for details.

\subsection{Computing unreliability}
\label{subsec:sft_unreliability}
Let $X$ by the random variable representing the failure of \BE $x$.
We use $\prob[t]{f_x} := \prob{X \leq t}$ to denote the probability that $x$ fails within time bound $t$.
A common metric is the \emph{unreliability} $\prob[t]{\dft} := \prob[t]{f_{\tle}}$, \ie the probability that the SFT $\dft$ with TLE $\tle$ fails within time bound $t$.
This metric can easily be computed on the BDD by employing Shannon decomposition: $\prob[t]{f} = \prob[t]{f_x} \cdot \prob[t]{\cof{f}{x=1}} + (1{-}\prob[t]{f_x}) \cdot \prob[t]{\cof{f}{x=0}}$.
The algorithm works independently of the calculation for $\prob[t]{f_x}$  and can therefore be applied to any failure distribution of the \BE{}s.

\paragraph{Computing sensitivity measures.}
Importance measures~\cite{DBLP:journals/csr/RuijtersS15,DBLP:books/cu/TrivediB17} are used to assess how sensitive the overall system is \wrt sub-systems.
The \emph{Birnbaum importance index}~\cite{birnbaum1969importance} is a prominent metric.
For \BE{} $e$ in SFT $\dft$ at time $t$, the Birnbaum index is given by the conditional probabilities $\BI_t(\dft, e) := \prob[t]{\dft \mid e} - \prob[t]{\dft \mid \neg e}$, where $\neg e$ represents that $e$ has not failed.
The calculation is done on the BDD corresponding to $\dft$, \cf\cite{DBLP:journals/ress/DutuitR01}.
Additional metrics are computed based on the Birnbaum index~\cite{DBLP:journals/ress/DutuitR01}, \eg the \emph{critical importance factor}, the \emph{Vesely-Fussell importance factor}~\cite{fussell1975hand}, the \emph{risk achievement worth} and the \emph{risk reduction worth}~\cite{cheok1998use}.

\paragraph{Vectorisation for multiple time bounds.}
Investigating the unreliability over time requires computing the unreliability for a large number of different time bounds.
The calculation for multiple time bounds can easily be parallelised, because the computations are independent:
the failure probability of the top event within time bound $t$ only depends on the failure probabilities of the \BE{}s within~$t$.
We employ \emph{vectorisation}~\cite{Eijkhout:IHPSClulu} where multiple probabilities---corresponding to different time bounds---are stored within a single vector and computed concurrently.
Vectorisation exploits both temporal and spatial locality in modern CPU caches as well as \emph{SIMD instructions} (single instruction multiple data) which operate on arrays of values at once.

\subsection{Computing the MTTF}
\label{subsec:mttf_approx}
Vectorisation naturally leads to approximation methods as we can efficiently evaluate many time bounds at once.
We exploit this for the \emph{mean-time-to-failure (MTTF)}, \ie the expected time point the SFT $\dft$ fails at.
It is calculated by $\int_0^\infty \prob[t]{\dft}\,\mathrm{d}t$.
We numerically approximate the improper integral by sampling a large number of time bounds on the BDD obtained for the SFT.
We use two different methods from~\cite{davis2007methods}.
The first method, \emph{proceeding to the limit}, computes a sequence of integrals $\int_{r_i}^{r_{i+1}} \prob[t]{\dft}\,\mathrm{d}t$ until the result of an integral is less than a given error $\varepsilon$.
Our implementation uses varying steps sizes which start at $10^{-10}$ and a default error of $\varepsilon := 10^{-12}$.
The second method, \emph{change of variable}, aims to ``squeeze'' the unbounded interval $[0,\infty)$ into the bounded interval $[0,1)$ using integration by substitution.
The latter method always uses the same number of samples (default $10^6$) and works good for functions slowly approaching zero.
The former method uses a variable number of samples and performs better for functions which approach zero relatively fast or change rapidly; see~\cite{Basgoeze20} for details.

\subsection{Implementation}
\label{subsec:sft_implementation}
We implemented the SFT analysis in the \stormdft\footnote{\url{https://www.stormchecker.org/}} tool based on the \storm model checker~\cite{hensel2021probabilistic} and use the multi-core library \sylvan~\cite{DBLP:journals/sttt/DijkP17} for creating and handling BDDs.
We list the main implementation details in the following.

\noindent\textbf{Multi-core BDD.} \tool{Sylvan} natively enables multi-core computations on BDD~\cite{DBLP:phd/basesearch/Dijk16}.
\stormdft exploits this when performing the translation from SFT to BDD.

\noindent\textbf{Complement edges.} The implementation uses \emph{complement edges}~\cite{DBLP:conf/dac/BraceRB90} which negate the corresponding function and it allows to use a single terminal node.

\noindent\textbf{Variable ordering.} The implementation uses the order of \BE{}s given in the input file (in Galileo format\footnote{\url{https://dftbenchmarks.utwente.nl/galileo.html}}) as the variable ordering for the BDD.
That way, we support arbitrary variables orderings which can be explicitly given by either the user or a pre-processing step.
Currently, \stormdft supports the DFS and TDLR variable orderings via pre-processing steps.

\noindent\textbf{Caching.} During the translation, intermediate BDDs are not cached in order to reduce the memory consumption.
Caching can be explicitly enabled for specific events if needed.

\noindent\textbf{Vectorisation.}
\stormdft uses the \tool{Eigen} library~\cite{eigenweb} for vectorisation.
The \emph{chunk-size}, \ie the number of time points computed in parallel, can be configured from the command-line.
We refer to~\cite{Basgoeze20} for details on the optimal chunk-size.

\noindent\textbf{Properties.} Apart from standard metrics presented before, our implementation supports properties defined in a fragment of \emph{continuous stochastic logic (CSL)}~\cite{DBLP:journals/tse/BaierHHK03}.
More precise, \stormdft supports (time-bounded) reachability formulas of the form $\mathbb{P}_{=?}(\lozenge^{\leq t} \phi)$ for state formula $\phi$ and time bound $t$.

%% file: figures/ex_sft.tex
\begin{tikzpicture}
	\node[and2] (T) at (8.0, -0.7) {};
	\node[labelbox] (T_label) at (T.east) {\underline{T}};
	\node[and3] (G) at (7.0, -2.3) {\rotatebox{270}{$2$}};
	\node[labelbox] (G_label) at (G.east) {G};
	\node[or2] (H) at (9.0, -2.3) {};
	\node[labelbox] (H_label) at (H.east) {H};
	\node[or2] (F) at (6.2, -4.0) {};
	\node[labelbox] (F_label) at (F.east) {F};
	\node[be] (A) at (5.7, -5.0) {};
	\node[labelbox] (A_label) at (A.north) {A};
	\node[be] (B) at (6.7, -5.0) {};
	\node[labelbox] (B_label) at (B.north) {B};
	\node[be] (C) at (7.4, -3.6) {};
	\node[labelbox] (C_label) at (C.north) {C};
	\node[be] (D) at (8.5, -3.6) {};
	\node[labelbox] (D_label) at (D.north) {D};
	\node[be] (E) at (9.5, -3.6) {};
	\node[labelbox] (E_label) at (E.north) {E};

	\draw[-](T.input 1) -- (G_label.north);
	\draw[-](T.input 2) -- (H_label.north);
	\draw[-](G.input 1) -- (F_label.north);
	\draw[-](G.input 2) -- (C_label.north);
	\draw[-](G.input 3) -- (D_label.north);
	\draw[-](H.input 1) -- (D_label.north);
	\draw[-](H.input 2) -- (E_label.north);
	\draw[-](F.input 1) -- (A_label.north);
	\draw[-](F.input 2) -- (B_label.north);
\end{tikzpicture}

%% file: figures/ex_bdd_dfs.tex
\begin{tikzpicture}[bdd]
    \node[innernode] at (0,0) (a) {$x_A$};
    \node[innernode] at (1.2,-1) (b) {$x_B$};
	\node[innernode] at (-1,-2) (c1) {$x_C$};
	\node[innernode] at (1.2,-2) (c2) {$x_C$};
	\node[innernode] at (-1,-3) (d1) {$x_D$};
	\node[innernode] at (0,-3) (d2) {$x_D$};
    \node[innernode] at (0.2,-4) (e) {$x_E$};
	\node[leaf] at (-1,-5.2) (one) {$1$};
	\node[leaf] at (1.2,-5.2) (zero) {$0$};

	\draw[succone]
	  (a) -- (c1)
	  (b) -- (c1)
	  (c1) -- (d1)
	  (c2) -- (d2)
	  (d1) -- (one.north)
      (d2) -- ($(one.north)!0.5!(one.north east)$)
	  (e) -- (one.north east)
	;
    \draw[succzero]
	  (a) -- (b)
	  (b) -- (c2)
	  (c1) -- (d2)
      (c2) -- (zero.north)
	  (d1) -- (e)
      (d2) .. controls +(-45:1.375) .. ($(zero.north)!0.5!(zero.north west)$)
	  (e) -- (zero.north west)
	;
\end{tikzpicture}

%% file: figures/ex_bdd_tdlr.tex
\begin{tikzpicture}[bdd]
	\node[innernode] at (0,0) (c) {$x_C$};
    \node[innernode] at (-1,-1) (d1) {$x_D$};
	\node[innernode] at (1.2,-1) (d2) {$x_D$};
    \node[innernode] at (-0.5,-2) (e) {$x_E$};
	\node[innernode] at (0,-3) (a) {$x_A$};
	\node[innernode] at (0.5,-4) (b) {$x_B$};
	\node[leaf] at (-1,-5.2) (one) {$1$};
	\node[leaf] at (1.2,-5.2) (zero) {$0$};

	\draw[succone]
	  (c) -- (d1)
	  (d1) -- (one.north)
	  (d2) -- (a)
	  (e) -- (a)
	  (a) -- ($(one.north)!0.5!(one.north east)$)
	  (b) -- (one.north east)
	;
    \draw[succzero]
	  (c) -- (d2)
	  (d1) -- (e)
      (d2) -- (zero.north)
      (e) .. controls +(-10:1) and ($(zero.north west)!0.250!(zero.north east)+(0,1.875)$) .. ($(zero.north west)!0.250!(zero.north east)$)
	  (a) -- (b)
	  (b) -- (zero.north west)
	;
\end{tikzpicture}

%% file: sections/evaluation_sft.tex
\section{Evaluation of SFT approach}
\label{sec:evaluation_sft}
We evaluate the fault tree analysis via BDDs as implemented in \stormdft on a range of benchmarks and compare with existing tools.
For reproducibility, we provide an artifact online\footnote{\url{https://doi.org/10.5281/zenodo.6390998}} which contains the analysis scripts, tool configurations and fault tree models used in our experimental evaluation.

\subsection{Configurations}
\label{subsec:configs}
We use \stormdft version 1.6.3.
In the default configuration, \stormdft is single-threaded, uses a chunk-size of 1024 for vectorisation and uses the DFS variable ordering.
We also evaluate \stormdft in a configuration using the TDLR variable ordering and in configurations using multiple cores.

We compare the SFT analysis in \stormdft with two existing tools: \scram\footnote{\url{https://github.com/rakhimov/scram}} (version 0.16.2) and \xfta\footnote{\url{https://altarica-association.org/members/arauzy/Software/XFTA/XFTA2.html}} (version 2.0.1).

\scram~\cite{scram18} is an open-source probabilistic risk analysis tool which supports the \emph{Open-PSA Model Exchange} format~\cite{openpsa}.
It performs SFT analysis using BDDs and supports metrics such as MCS, unreliability and importance measures.
Before analysis, \scram simplifies the SFT's graph structure.
The BDD variable ordering then follows the topological ordering on the simplified SFT.

\xfta~\cite{rauzy2020probabilistic} is a free-to-use tool for the analysis of fault trees and similar models.
It is hosted by the AltaRica Association.
\xfta uses its own object-oriented design language \emph{S2ML+SBE} as input, but also supports the Open-PSA format.
The analysis is performed by either generating the MCS and calculating the metrics on them or by creating a BDD and computing the metrics via the BDD.
For the variable ordering, the children of gates are sorted beforehand and then the DFS ordering is used.

\subsection{Benchmarks}
\label{subsec:benchmarks}
We use the following collection of SFT benchmarks for our evaluation:
\begin{compactitem}
    \item 40 examples from the Aralia benchmark set\footnote{\url{https://github.com/rakhimov/scram/tree/develop/input/Aralia}}. We excluded 3 non-coherent SFTs containing a negation-gate as \stormdft does not support them.
    \item 3 models of wet-pipe fire sprinkler systems in Australian shopping centres~\cite{moinuddin2019reliability}
    \item 8 examples modelling train routing options \wrt infrastructure failures in railway station areas~\cite{DBLP:conf/fmics/0001WKN19}.
    \item 3 industrial models for components of a lock used in water navigation.
    \item 161 randomly generated SFTs using a script provided by the \scram tool.
        128 of the random SFTs have 150 \BE{}s and 33 are large SFTs with 500 \BE{}s.
\end{compactitem}

\begin{table}[t]
    \centering
    \caption{SFT benchmark sizes}
    \begin{tabular}{l|cccccc}
        \toprule
        & Aralia & Sprinkler & Railway & Industry & Random & Random (Large) \\
        \midrule
        \#BEs & 25--1567 & 31 & 22--54 & 36--184 & 150 & 500\\
        \#Gates & 20--1622 & 35 & 69--259 & 21--67 & 70-122& 261-316\\
        \bottomrule
    \end{tabular}
    \label{tab:benchmark_sizes_sft}
\end{table}
We provide statistics on the benchmarks in Table~\ref{tab:benchmark_sizes_sft}.
We give the minimal and maximal number of \BE{}s and gates for each benchmark set.

We analyse each fault tree \wrt four queries: all minimal cut sets (\emph{MCS}), the \emph{unreliability} at time point $t=1$, the \emph{unreliability for \num{10000} time points} that are uniformly distributed within the interval $[0, 10]$, and the \emph{Birnbaum importance index} at time point $t=1$.

\subsection{Results}
\label{subsec:results}
We ran \stormdft, \scram and \xfta on all 215 examples \wrt the four different queries.
We ran the experiments on a desktop machine with an AMD Ryzen™ 9 5950X and \SI{32}{\giga\byte} of RAM running Arch Linux.
In the multi-core configuration, we used \num{16} cores.
The timeout was set to \SI{5}{\minute} and the memory was limited to \SI{30}{\giga\byte}.
We asserted that the obtained results are the same for all three tools.
In the following, we provide detailed comparisons of the tools.
Additional results and details can be found in
\ifextended
App.~\ref{app:additional_results}.
\else
the extended version of this paper~\cite{DBLP:journals/corr/abs-2202-02829}.
\fi

\begin{figure}[t]
    \centering
    \subfigure[Runtime \stormdft vs \scram]{
        \centering
        \makebox[0.47\linewidth]{
        \input{figures/sft_plots/storm_scram_mcs}}
        \label{fig:storm_scram_mcs}
    }
    \subfigure[Runtime \stormdft vs \xfta]{
        \centering
        \makebox[0.47\linewidth]{
        \input{figures/sft_plots/storm_xfta_mcs}}
        \label{fig:storm_xfta_mcs}
    }
    \subfigure[Runtime \stormdft using variable orderings DFS vs TDLR]{
        \centering
        \makebox[0.47\linewidth]{
        \input{figures/sft_plots/storm_dfs_tdlr}}
        \label{fig:storm_dfs_tdlr}
    }
    \subfigure[Runtime single- vs multi-core]{
        \makebox[0.47\linewidth]{
        \input{figures/sft_plots/storm_mult_storm.tex}}
        \label{fig:storm_mult_storm_mcs}
    }
    \caption{Comparisons for the computation of MCS}
    \label{fig:eval_mcs}
\end{figure}
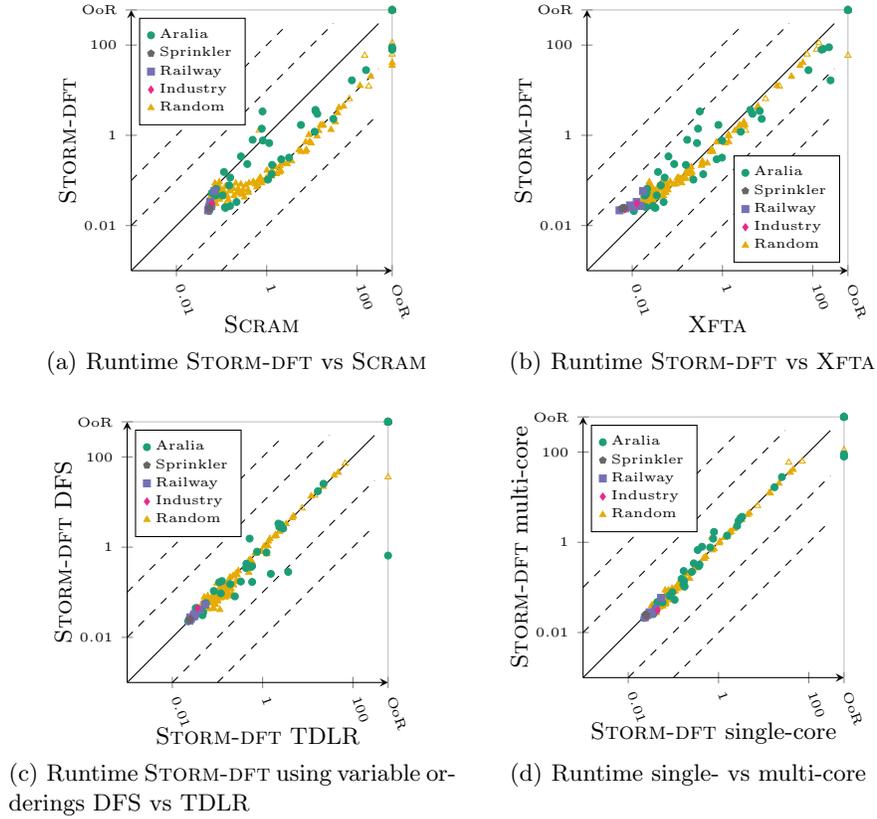

We present the comparisons as scatter plots such as in Fig.~\ref{fig:eval_mcs}.
All scatter plots are in log-log scale and indicate---in most cases---the time (in seconds) it took each tool to compute a query.
Line \emph{OoR} indicates \emph{out of resources} and represents either a timeout or memory out.
All points below the diagonal indicate examples which \stormdft could solve faster than the other tool.
All points below the first (second) dashed line correspond to SFTs for which \stormdft was one (two) order(s) of magnitude faster than the other tool.
Similarly, for every point above the diagonal, the other tool was faster.

\bfpar{BDD sizes.}
\begin{figure}[t]
    \centering
    \subfigure[BDD sizes \stormdft DFS vs \scram]{
        \centering
        \makebox[0.47\linewidth]{
        \input{figures/sft_plots/bdd_sizes_storm_scram}}
        \label{fig:bdd_sizes_storm_scram}
    }
    \subfigure[BDD sizes \stormdft using variable ordering from \scram vs \scram]{
        \makebox[0.47\linewidth]{
        \input{figures/sft_plots/bdd_sizes_scram.tex}}
        \label{fig:bdd_sizes_scram}
    }
    \caption{Comparison of BDD sizes for different variable orderings for MCS}
    \label{fig:bdd_sizes_mcs}
\end{figure}
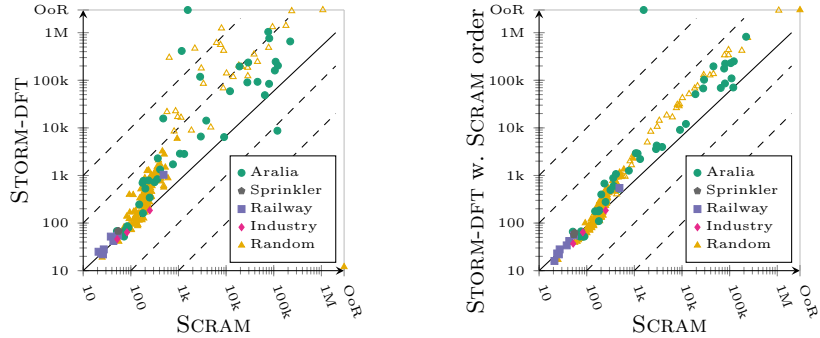
First, Fig.~\ref{fig:bdd_sizes_mcs} compares the number of nodes in the BDDs which provides an idea of the respective memory consumption.
Fig.~\ref{fig:bdd_sizes_storm_scram} compares the sizes of the BDDs obtained by \stormdft and \scram.
The largest BDDs which could be analysed contain more than a million nodes.
In general, \scram yields smaller BDDs, for larger BDDs even by more than one order of magnitude.
The main reason is that \scram uses a slightly different variable ordering which seems to yield smaller BDDs.

The influence of the variable ordering is further investigated in Fig.~\ref{fig:bdd_sizes_scram}.
Here, we extract the variable ordering from \scram and employ it in \stormdft.
We see that using the \scram variable ordering in \stormdft improves upon the default DFS ordering and yields smaller BDD sizes---in particular for larger SFTs.
However, \scram still yields smaller BDDs than \stormdft.
Reasons for the discrepancy could be that the BDD implementation in \scram is specifically tailored to SFTs or that the self-reported number of BDD nodes in both tools are computed in different ways.

\noindent\emph{\textcolor{blue}{
Using good heuristics for the variable ordering is crucial for a small memory footprint; the heuristics in \stormdft can be further improved.}}

\bfpar{MCS.}
We compare the runtimes for computing the MCS in Fig.~\ref{fig:eval_mcs}.
Fig.~\ref{fig:storm_scram_mcs} compares \stormdft and \scram.
We first see that the SFTs corresponding to the sprinkler, railway and industry case studies can be solved within \SI{1}{\second} by both tools.
This also holds for other tools/configurations and metrics.
As these SFTs are not a challenge, we focus on the Aralia benchmark and random SFTs in the remainder.
\stormdft is faster than \scram in nearly all cases.
One possible reason is that \scram outputs the MCS in an XML format which requires more I/O-operations than the simple list output of \stormdft.

When comparing \stormdft with \xfta (\cf Fig.~\ref{fig:storm_xfta_mcs}), the picture is more diverse.
\xfta is faster than \stormdft on most examples which can be solved within \SI{1}{\second}.
This is mostly due to the overhead resulting from initializing the \sylvan BDD library within \storm.
For some of the medium-sized Aralia examples, \xfta performs better than \stormdft.
However, for larger examples, \stormdft prevails on all the random SFTs and nearly all Aralia benchmarks.

\noindent\emph{\textcolor{blue}{For MCS, \stormdft is faster than both \scram and \xfta for larger SFTs.}}

Fig.~\ref{fig:storm_dfs_tdlr} compares the runtimes of \stormdft for different variable orderings DFS and TDLR.
While both variable orderings do not make much of a difference for most examples, DFS performs better for some of the fault trees and even allows to handle an FT which is OoR for TDLR.

Fig.~\ref{fig:storm_mult_storm_mcs} shows that using \num{16} cores (instead of a single core) for the BDD operations as supported by the \sylvan library has only a minor influence on the runtime.
One reason is that most operations performed on the BDDs are fairly basic and therefore do not profit much from parallelization.
However, for some large examples, the configuration with multiple cores allows to handle SFTs which were OoR for the single core.

\bfpar{Unreliability.}
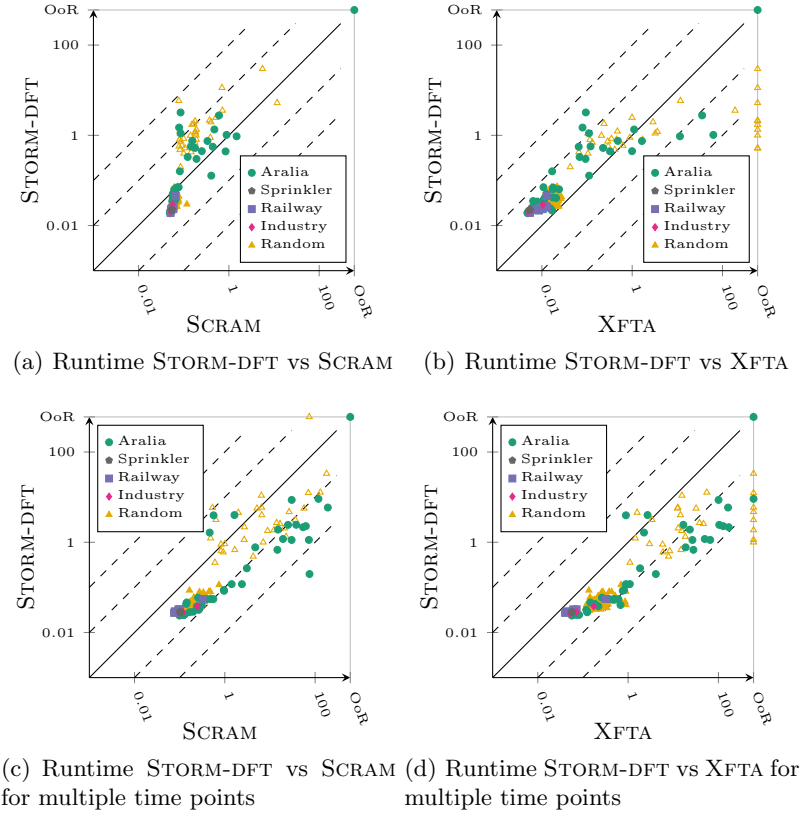
\begin{figure}[t]
    \centering
    \subfigure[Runtime \stormdft vs \scram]{
        \centering
        \input{figures/sft_plots/storm_scram_prob}
        \label{fig:storm_scram_prob}
    }
    \subfigure[Runtime \stormdft vs \xfta]{
        \centering
        \input{figures/sft_plots/storm_xfta_prob}
        \label{fig:storm_xfta_prob}
    }
    \subfigure[Runtime \stormdft vs \scram for multiple time points]{
        \centering
        \input{figures/sft_plots/storm_scram_prob_vec}
        \label{fig:storm_scram_prob_vec}
    }
    \subfigure[Runtime \stormdft vs \xfta for multiple time points]{
        \centering
        \input{figures/sft_plots/storm_xfta_prob_vec}
        \label{fig:storm_xfta_prob_vec}
    }
    \caption{Comparisons for the computation of the unreliability}
    \label{fig:eval_unrel}
\end{figure}
Fig.~\ref{fig:eval_unrel} shows the runtimes for computing the unreliability.
Fig.~\ref{fig:storm_scram_prob} compares \stormdft and \scram.
For most examples, both tools compute the unreliability within \SI{1}{\second}.
For larger benchmarks, \scram outperforms  \stormdft.
The main reason is that \stormdft builds larger BDDs than \scram, cf Fig.~\ref{fig:bdd_sizes_storm_scram}.
\xfta is faster than \stormdft on the small examples, \cf Fig.~\ref{fig:storm_xfta_prob}.
However, for larger SFTs, \stormdft outperforms \xfta.
For all tools, computing the unreliability is significantly faster than computing the MCS.

Fig.~\ref{fig:storm_scram_prob_vec} compares the performance of \stormdft and \scram when computing the unreliability for \num{10000} different time points.
\stormdft performs vectorisation with a chunk-size of \num{1024} and thus computes \num{1024} time points at once.
This dedicated support yields a clear performance gain compared to \scram which computes each time point sequentially.
For larger SFTs, \stormdft is more than one order of magnitude faster.
The same holds true when comparing to \xfta, \cf Fig.~\ref{fig:storm_xfta_prob_vec}.

\noindent\emph{\textcolor{blue}{\stormdft is slower when computing the unreliability for one time bound, but is significantly faster than \scram and \xfta for multiple time bounds.}}

\bfpar{Importance measures.}
\begin{figure}[t]
    \centering
    \subfigure[Runtime \stormdft vs \xfta for single time point]{
        \input{figures/sft_plots/storm_xfta_importance}
        \centering
        \label{fig:storm_xfta_importance}
    }
    \subfigure[Runtime \stormdft vs \xfta for multiple time points]{
        \centering
        \input{figures/sft_plots/storm_xfta_importance_vec}
        \label{fig:storm_xfta_importance_vec}
    }
    \caption{Comparison for computing the Birnbaum importance index for all \BE{}s}
    \label{fig:eval_importance}
\end{figure}
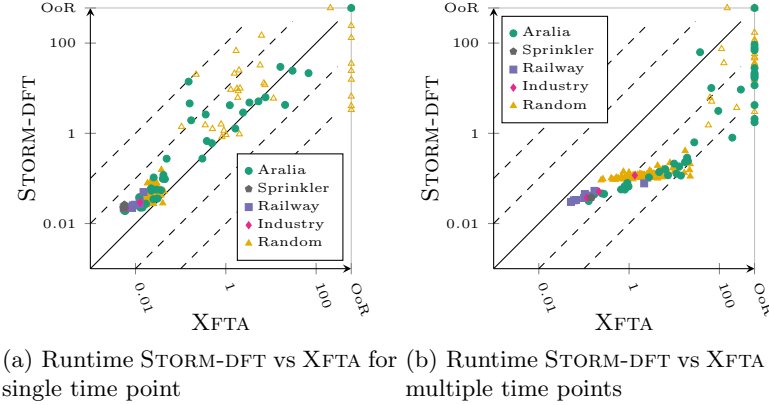
Last, we consider the computation of the Birnbaum importance index for all \BE{}s in a SFT.
We omit the results for \scram as \scram needs to compute the MCS for computing the Birnbaum importance index for all \BE{}s.
As \stormdft does not need this computation, the comparison would be unfair.
We provide the results in
\ifextended
App.~\ref{app:additional_results}.
\else
the extended version of this paper~\cite{DBLP:journals/corr/abs-2202-02829}.
\fi

Fig.~\ref{fig:storm_xfta_importance} compares the runtime for \stormdft and \xfta when computing the Birnbaum importance index for all \BE{}s at a single time point.
We see that most examples are solved within \SI{1}{\second}.
\xfta performs better on the larger examples.
Fig.~\ref{fig:storm_xfta_importance_vec} compares both tools when computing the Birnbaum importance index for \num{1000} time points.
Here, \stormdft is orders of magnitude faster than \xfta and provides results where \xfta runs out of resources.

\noindent\emph{\textcolor{blue}{When computing the Birnbaum importance index for all \BE{}s, \xfta is faster for single time points whereas \stormdft is orders of magnitude faster for multiple time points.}}

%% file: figures/sft_plots/bdd_sizes_storm_scram.tex
\begin{tikzpicture}
  \begin{axis}[scatterplot,
    xmin=10, ymin=10, ymax=3000000, xmax=3000000, xmode=log, ymode=log,
    xtick={10, 100, 1000, 10000, 100000, 1000000}, xticklabels={10, 100, 1k, 10k, 100k, 1M},
    extra x ticks = {3000000}, extra x tick labels = {OoR}, extra x tick style = {grid = major},
    ytick={10, 100, 1000, 10000, 100000, 1000000}, yticklabels={10, 100, 1k, 10k, 100k, 1M},
    extra y ticks = {3000000}, extra y tick labels = {OoR}, extra y tick style = {grid = major},
    xlabel=\scram, ylabel=\stormdft,
    legend pos=south east]
  \addplot[
    scatter,only marks,
    scatter/classes={
      aralia={aralia},
      sprinkler={sprinkler},
      railway={railway},
      industry={industry},
      randomScram={randomScram},
      randomScramProb={randomScramProb}
    },
    scatter src=explicit symbolic]
    table [col sep=semicolon, x=BDD-SIZE-SCRAM, y=BDD-SIZE-STORM-DFS, meta=Set]
    {figures/sft_plots/sft_sizes.csv};
  \legend{Aralia,Sprinkler,Railway,Industry,Random}
  \addplot[no marks] coordinates
    {(10,10) (2000000,1000000) };
  \addplot[no marks, dashed] coordinates
    {(100,10) (2000000,200000) };
  \addplot[no marks, dashed] coordinates
    {(1000,10) (2000000,20000) };
  \addplot[no marks, dashed] coordinates
    {(10,100) (200000,2000000) };
  \addplot[no marks, dashed] coordinates
    {(10,1000) (20000,2000000) };
  \end{axis}
\end{tikzpicture}

%% file: figures/sft_plots/bdd_sizes_scram.tex
\begin{tikzpicture}
  \begin{axis}[scatterplot,
    xmin=10, ymin=10, ymax=3000000, xmax=3000000, xmode=log, ymode=log,
    xtick={10, 100, 1000, 10000, 100000, 1000000}, xticklabels={10, 100, 1k, 10k, 100k, 1M},
    extra x ticks = {3000000}, extra x tick labels = {OoR}, extra x tick style = {grid = major},
    ytick={10, 100, 1000, 10000, 100000, 1000000}, yticklabels={10, 100, 1k, 10k, 100k, 1M},
    extra y ticks = {3000000}, extra y tick labels = {OoR}, extra y tick style = {grid = major},
    xlabel=\scram, ylabel=\stormdft w. \scram order,
    y label style={at={(axis description cs:-0.18,0.42)},anchor=south},
    legend pos=south east]
  \addplot[
    scatter,only marks,
    scatter/classes={
      aralia={aralia},
      sprinkler={sprinkler},
      railway={railway},
      industry={industry},
      randomScram={randomScram},
      randomScramProb={randomScramProb}
    },
    scatter src=explicit symbolic]
    table [col sep=semicolon, x=BDD-SIZE-SCRAM, y=BDD-SIZE-STORM-SCRAM, meta=Set]
    {figures/sft_plots/sft_sizes.csv};
  \legend{Aralia,Sprinkler,Railway,Industry,Random}
  \addplot[no marks] coordinates
    {(10,10) (2000000,1000000) };
  \addplot[no marks, dashed] coordinates
    {(100,10) (2000000,200000) };
  \addplot[no marks, dashed] coordinates
    {(1000,10) (2000000,20000) };
  \addplot[no marks, dashed] coordinates
    {(10,100) (200000,2000000) };
  \addplot[no marks, dashed] coordinates
    {(10,1000) (20000,2000000) };
  \end{axis}
\end{tikzpicture}

%% file: figures/sft_plots/storm_scram_mcs.tex
\begin{tikzpicture}
  \begin{axis}[scatterruntime,
      xmode=log, ymode=log,
      xlabel=\scram, ylabel=\stormdft,
    ]
  \addplot[
    scatter,only marks,
    scatter/classes={
      aralia={aralia},
      sprinkler={sprinkler},
      railway={railway},
      industry={industry},
      randomScram={randomScram},
      randomScramProb={randomScramProb}
    },
    scatter src=explicit symbolic]
    table [col sep=semicolon, x=SCRAM_MCS, y=STORM_MCS_DFS, meta=Set]
    {figures/sft_plots/sft_plots.csv};
  \legend{Aralia,Sprinkler,Railway,Industry,Random}
  \diagonalsTimes
  \end{axis}
\end{tikzpicture}

%% file: figures/sft_plots/storm_xfta_mcs.tex
\begin{tikzpicture}
  \begin{axis}[scatterruntime,
      xmode=log, ymode=log,
      xlabel=\xfta, ylabel=\stormdft,
      legend pos=south east
    ]
  \addplot[
    scatter,only marks,
    scatter/classes={
      aralia={aralia},
      sprinkler={sprinkler},
      railway={railway},
      industry={industry},
      randomScram={randomScram},
      randomScramProb={randomScramProb}
    },
    scatter src=explicit symbolic]
    table [col sep=semicolon, x=XFTA_MCS, y=STORM_MCS_DFS, meta=Set]
    {figures/sft_plots/sft_plots.csv};
  \legend{Aralia,Sprinkler,Railway,Industry,Random}
  \diagonalsTimes
  \end{axis}
\end{tikzpicture}

%% file: figures/sft_plots/storm_dfs_tdlr.tex
\begin{tikzpicture}
  \begin{axis}[scatterruntime,
      xmode=log, ymode=log,
      xlabel=\stormdft TDLR, ylabel=\stormdft DFS,
    ]
  \addplot[
    scatter,only marks,
    scatter/classes={
      aralia={aralia},
      sprinkler={sprinkler},
      railway={railway},
      industry={industry},
      randomScram={randomScram},
      randomScramProb={randomScramProb}
    },
    scatter src=explicit symbolic]
    table [col sep=semicolon, x=STORM_MULT_MCS_TDLR, y=STORM_MULT_MCS_DFS, meta=Set]
    {figures/sft_plots/sft_plots.csv};
  \legend{Aralia,Sprinkler,Railway,Industry,Random}
  \diagonalsTimes
  \end{axis}
\end{tikzpicture}

%% file: figures/sft_plots/storm_mult_storm.tex
\begin{tikzpicture}
  \begin{axis}[scatterruntime,
      xmode=log, ymode=log,
      xlabel=\stormdft single-core, ylabel=\stormdft multi-core,
    ]
  \addplot[
    scatter,only marks,
    scatter/classes={
      aralia={aralia},
      sprinkler={sprinkler},
      railway={railway},
      industry={industry},
      randomScram={randomScram},
      randomScramProb={randomScramProb}
    },
    scatter src=explicit symbolic]
    table [col sep=semicolon, x=STORM_MULT_MCS_DFS, y=STORM_MCS_DFS, meta=Set]
    {figures/sft_plots/sft_plots.csv};
  \legend{Aralia,Sprinkler,Railway,Industry,Random}
  \diagonalsTimes
  \end{axis}
\end{tikzpicture}

%% file: figures/sft_plots/storm_scram_prob.tex
\begin{tikzpicture}
  \begin{axis}[scatterruntime,
    xmode=log, ymode=log,
    xlabel=\scram, ylabel=\stormdft,
    legend pos=south east]
  \addplot[
    scatter,only marks,
    scatter/classes={
      aralia={aralia},
      sprinkler={sprinkler},
      railway={railway},
      industry={industry},
      randomScram={randomScram},
      randomScramProb={randomScramProb}
    },
    scatter src=explicit symbolic]
    table [col sep=semicolon, x=SCRAM_PROB, y=STORM_PROB_DFS, meta=Set]
    {figures/sft_plots/sft_plots.csv};
  \legend{Aralia,Sprinkler,Railway,Industry,Random}
  \diagonalsTimes
  \end{axis}
\end{tikzpicture}

%% file: figures/sft_plots/storm_xfta_prob.tex
\begin{tikzpicture}
  \begin{axis}[scatterruntime,
    xmode=log, ymode=log,
    xlabel=\xfta, ylabel=\stormdft,
    legend pos=south east]
  \addplot[
    scatter,only marks,
    scatter/classes={
      aralia={aralia},
      sprinkler={sprinkler},
      railway={railway},
      industry={industry},
      randomScram={randomScram},
      randomScramProb={randomScramProb}
    },
    scatter src=explicit symbolic]
    table [col sep=semicolon, x=XFTA_PROB, y=STORM_PROB_DFS, meta=Set]
    {figures/sft_plots/sft_plots.csv};
  \legend{Aralia,Sprinkler,Railway,Industry,Random}
  \diagonalsTimes
  \end{axis}
\end{tikzpicture}

%% file: figures/sft_plots/storm_scram_prob_vec.tex
\begin{tikzpicture}
  \begin{axis}[scatterruntime,
    xmode=log, ymode=log,
    xlabel=\scram, ylabel=\stormdft]
  \addplot[
    scatter,only marks,
    scatter/classes={
      aralia={aralia},
      sprinkler={sprinkler},
      railway={railway},
      industry={industry},
      randomScram={randomScram},
      randomScramProb={randomScramProb}
    },
    scatter src=explicit symbolic]
    table [col sep=semicolon, x=SCRAM_PROB_VEC, y=STORM_PROB_DFS_VEC, meta=Set]
    {figures/sft_plots/sft_plots.csv};
  \legend{Aralia,Sprinkler,Railway,Industry,Random}
  \diagonalsTimes
  \end{axis}
\end{tikzpicture}

%% file: figures/sft_plots/storm_xfta_prob_vec.tex
\begin{tikzpicture}
  \begin{axis}[scatterruntime,
    xmode=log, ymode=log,
    xlabel=\xfta, ylabel=\stormdft]
  \addplot[
    scatter,only marks,
    scatter/classes={
      aralia={aralia},
      sprinkler={sprinkler},
      railway={railway},
      industry={industry},
      randomScram={randomScram},
      randomScramProb={randomScramProb}
    },
    scatter src=explicit symbolic]
    table [col sep=semicolon, x=XFTA_PROB_VEC, y=STORM_PROB_DFS_VEC, meta=Set]
    {figures/sft_plots/sft_plots.csv};
  \legend{Aralia,Sprinkler,Railway,Industry,Random}
  \diagonalsTimes
  \end{axis}
\end{tikzpicture}

%% file: figures/sft_plots/storm_xfta_importance.tex
\begin{tikzpicture}
  \begin{axis}[scatterruntime,
      xmode=log, ymode=log,
      xlabel=\xfta, ylabel=\stormdft,
      legend pos=south east
    ]
  \addplot[
    scatter,only marks,
    scatter/classes={
      aralia={aralia},
      sprinkler={sprinkler},
      railway={railway},
      industry={industry},
      randomScram={randomScram},
      randomScramProb={randomScramProb}
    },
    scatter src=explicit symbolic]
    table [col sep=semicolon, x=XFTA_IMP, y=STORM_IMP, meta=Set]
    {figures/sft_plots/sft_plots.csv};
  \legend{Aralia,Sprinkler,Railway,Industry,Random}
  \diagonalsTimes
  \end{axis}
\end{tikzpicture}

%% file: figures/sft_plots/storm_xfta_importance_vec.tex
\begin{tikzpicture}
  \begin{axis}[scatterruntime,
      xmode=log, ymode=log,
      xlabel=\xfta, ylabel=\stormdft,
    ]
  \addplot[
    scatter,only marks,
    scatter/classes={
      aralia={aralia},
      sprinkler={sprinkler},
      railway={railway},
      industry={industry},
      randomScram={randomScram},
      randomScramProb={randomScramProb}
    },
    scatter src=explicit symbolic]
    table [col sep=semicolon, x=XFTA_IMP_VEC, y=STORM_IMP_VEC, meta=Set]
    {figures/sft_plots/sft_plots.csv};
  \legend{Aralia,Sprinkler,Railway,Industry,Random}
  \diagonalsTimes
  \end{axis}
\end{tikzpicture}

%% file: sections/dft-bdd.tex
\section{DFT analysis via BDD and modularisation}
\label{sec:dft-bdd}

Dynamic fault trees extend SFTs by capturing dynamic failure behaviour such as ordered failures, spare management or functional dependencies.
Analysis of DFTs therefore needs to keep track of the history of failures and BDDs cannot be easily used.
In this approach, we combine SFT and DFT analysis using \emph{modularisation}~\cite{gulati1997modular}.
Modularisation is a ``divide-and-conquer''-approach which splits the DFT into \emph{modules}, \ie independent sub-trees.
Each module is analysed independently and the corresponding results are combined in the end.
Modularisation thus allows to exploit the ``best'' analysis technique for each module individually.

Modules containing dynamic elements are analysed by translating the corresponding sub-DFT into a Markov model~\cite{DBLP:journals/tii/VolkJK18}.
The state space of the Markov model is created by exhaustively exploring all possible \BE failures of the DFT.
Each transition corresponds to the failure of a \BE and the successor state represents the status of the DFT after the \BE failure.
The transition rate is given by the failure rate of the \BE.
Our translation employs several optimisation techniques to mitigate a state space explosion.
The optimisations encompass discarding irrelevant failures and exploiting symmetries, see~\cite{DBLP:journals/tii/VolkJK18} for the details.

Note that modularisation can only be used for computing probabilities, \eg the unreliability.
The MTTF cannot be computed compositionally as combining expectations is difficult~\cite{DBLP:conf/papm/BohnenkampH02}.
Thus, other approaches are necessary such as the approximation from Sect.~\ref{subsec:mttf_approx} or composing independent Markov models~\cite{DBLP:journals/tii/VolkJK18}.

\paragraph{Algorithm.}
\begin{algorithm}[t]
    \caption{DFT analysis via modularisation}
    \label{alg:modularisation}
    \begin{algorithmic}
        \Require DFT $\dft$, time bounds $t_1, \dots, t_n$
        \Ensure Analysis results $\prob[t_1]{\dft}, \dots, \prob[t_n]{\dft}$
        \State Compute the modules $D = \set{\dft_1, \dots, \dft_k}$ in $\dft$
        \For{$\dft_i \in D$ sorted by decreasing size of $\dft_i$}
            \If{$\dft_i \setminus \bigcup_{\dft' \neq \dft_i} \dft'$ contains no dynamic gate}
                \State $D := D \setminus \dft_i$
            \EndIf
        \EndFor
        \For{$\dft_i \in D$}
            \State Generate Markov model $\ctmc_i$ from $\dft_i$
            \State Compute failure probabilities $p_1 = \prob[t_1]{\ctmc_i}, \dots, p_n = \prob[t_n]{\ctmc_i}$ on $\ctmc_i$
            \State Create \BE $B_i$ such that $\prob[t_j]{B_i} = p_j$ for all $1 \leq j \leq n$
            \State Replace $\dft_i$ by $B_i$ in $\dft$
        \EndFor
        \State Build BDD $\bdd$ from $\dft$
        \State Compute results $r_1 = \prob[t_1]{\bdd}, \dots, r_n = \prob[t_n]{\bdd}$ on $\bdd$
        \State \Return $r_1, \dots, r_n$
    \end{algorithmic}
\end{algorithm}
We shortly describe our implementation of the DFT analysis via modularisation based on~\cite{gulati1997modular}.
Alg.~\ref{alg:modularisation} presents the pseudo-code.
We use the DFT in Fig.~\ref{fig:ex_dft_modular} as an example and compute the unreliability within time bound $t$.
We start the analysis by identifying the modules in the DFT using the algorithm of~\cite{DBLP:journals/tr/DutuitR96}.
The algorithm traverses the fault tree in a depth-first left-most order and stores the order in which nodes are visited.
A node $v$ is a root of a module if all its descendants are visited in-between the first and last visit of $v$.
The algorithm runs in linear time and yields a unique list of modules.
Minor adaptions of the algorithm are required to adequately handle \SEQ and \FDEP, \cf\cite{Basgoeze20}.
Next, we only keep the dynamic modules, \ie modules containing at least one dynamic element.
We iteratively remove a module if its corresponding elements not contained in other modules only contain static gates.
That way, we remove modules containing only static elements and modules which are a subset of dynamic modules.
This step results in a unique set of dynamic modules.
The example DFT contains two dynamic modules (indicated by dashed blue boxes on the left DFT in Fig.~\ref{fig:ex_dft_modular}).
Next, each dynamic module is translated to a Markov model and analysed according to the given metric~\cite{DBLP:journals/tii/VolkJK18} .
The complete dynamic module is then replaced by a single \BE which matches the calculated failure probabilities.
In our example, \BE $H'$ is chosen such that it has the same probability to fail within time bound $t$ than the whole module of $H$.
In the end, the resulting fault tree (on the right in Fig.~\ref{fig:ex_dft_modular}) contains only static elements.
Thus, this SFT can be analysed using the BDD approach presented in Sect.~\ref{sec:sft-bdd}.

Static modules could of course also be replaced by corresponding \BE{}s.
As building the BDD is efficient, we opt to directly analyse the resulting SFT instead.
Specific dynamic structures such as the first child of a \PAND or \SEQ could also be further modularised following~\cite{yevkin2011improved}.
However, the application of these modularisation rules is very limited and results in semi-Markov chains.
This approach is therefore not considered here.

\paragraph{Implementation.}
\label{subsec:dft_implementation}
We implemented the modularisation in \stormdft using the BDD implementation described in Sect.~\ref{subsec:sft_implementation}.
A DFT is analysed by translating it into a Markov model as in~\cite{DBLP:journals/tii/VolkJK18}.
While \stormdft already supports a modularisation, this top-down approach is only applicable to children of the top-level event.
In contrast, the new implementation is applicable to dynamic modules located anywhere in the DFT.
Moreover, the Markov models for dynamic modules are cached such that multiple queries can be performed on the same model.
This is not possible in the previous implementation which regenerates each model for a new metric.
The caching is in particular useful when computing multiple time points and exploiting vectorisation on the resulting SFT.

%% file: sections/evaluation_dft.tex
\section{Evaluation of DFT approach}
\label{sec:evaluation_dft}
We evaluate the DFT modularisation and compare with existing approaches.

\paragraph{Configurations.}
We compare the modularisation using BDDs with two existing approaches within \stormdft~\cite{DBLP:journals/tii/VolkJK18}: the translation to a continuous-time Markov chain (CTMC) and the top-down modularisation.

\paragraph{Benchmarks.}
We use the following DFT benchmarks:
\begin{compactitem}
    \item 68 DFTs from the \ffort benchmark collection~\cite{ruijters2019ffort}.
    \item 40 DFT obtained by using the SFTs from the Aralia benchmark set and replacing one \BE by the DFT \texttt{ftpp.1-1} from the \ffort benchmark set.

    \item 8 DFTs modelling infrastructure in railway station areas~\cite{DBLP:conf/fmics/0001WKN19} and slightly adapted to contain modules.
    \item 8 DFTs modelling configurations for a vehicle guidance system (VGS)~\cite{DBLP:journals/ress/GhadhabJKKV19} and adapted by removing irrelevant \FDEP{}s.
\end{compactitem}
Table~\ref{tab:benchmark_sizes_dft} gives  statistics on these benchmarks: the minimal and maximal number of \BE{}s, static and dynamic gates in the original DFT as well as the numbers for the SFT after modularisation.

\begin{table}[t]
    \centering
    \caption{DFT benchmark sizes}
    \begin{tabular}{l|ccc|cc}
        \toprule
        Benchmark set  & \#BEs & \#Static gates & \#Dyn. gates & \#BEs mod. & \#Static gates mod.\\
        \midrule
        Adapt. SFT     & 32-1574 & 26-1628 & 3     & 25-1623 & 21-1623\\
        Adapt. Railway & 194-545 & 153-487 & 19-54 & 22-54   & 40-168\\
        Adapt. VGS     & 54-99   & 31-59   & 6-20  & 1-79    & 0-39\\
        \ffort         & 6-87    & 1-50    & 0-44  & 1-50    & 0-21\\
        \bottomrule
    \end{tabular}
    \label{tab:benchmark_sizes_dft}
\end{table}

We run the three configurations of \stormdft on the 124 DFTs.
We compute the \emph{unreliability} at a time bound $t$ either given by the largest bound specified in \ffort or we use $t=100$ otherwise.
For \emph{multiple time bounds}, we use \num{1000} time bounds uniformly distributed over the interval $[0, t]$.
We used the same machine and settings (timeout \SI{5}{\minute}, \SI{30}{\giga\byte} memory) as in Sect.~\ref{subsec:configs}.

\subsection{Results}
\label{subsec:results_dft}

\bfpar{Unreliability.}
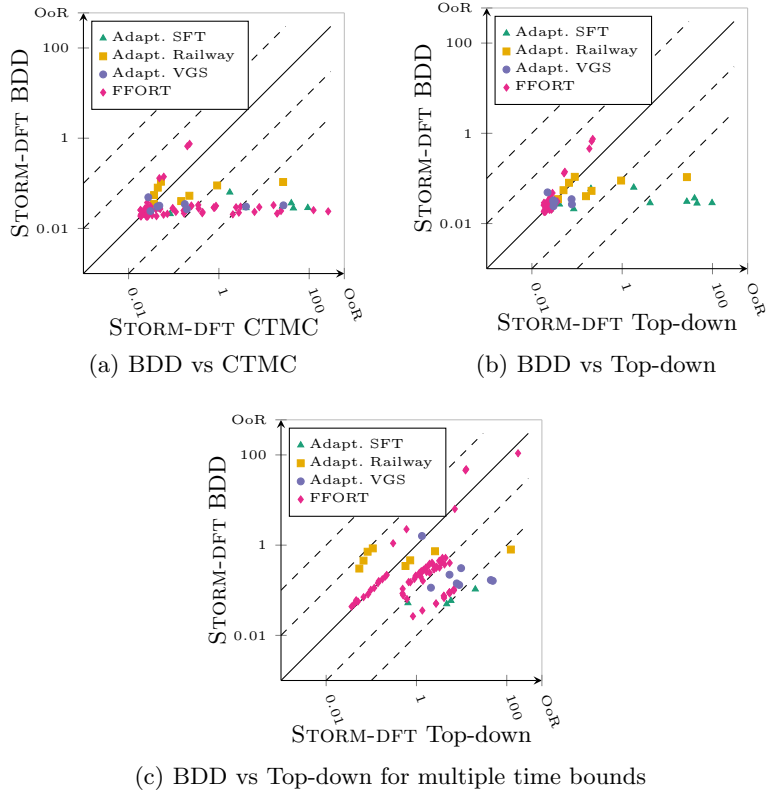
\begin{figure}[t]
    \centering
    \subfigure[BDD vs CTMC]{
        \centering
        \input{figures/dft_plots/dft_bdd_ctmc}
        \label{fig:dft_bdd_ctmc}
    }
    \subfigure[BDD vs Top-down]{
        \centering
        \input{figures/dft_plots/dft_bdd_ctmcmod}
        \label{fig:dft_bdd_ctmcmod}
    }
    \subfigure[BDD vs Top-down for multiple time bounds]{
        \makebox[0.6\linewidth]{
        \centering
        \input{figures/dft_plots/dft_bdd_vec_ctmcmod_vec}
        }
        \label{fig:dft_bdd_vec_ctmcmod_vec}
    }
    \caption{Comparisons of runtimes for the computation of the unreliability}
    \label{fig:eval_dft_rel}
\end{figure}
Fig.~\ref{fig:eval_dft_rel} compares the computation of the unreliability.
Fig.~\ref{fig:dft_bdd_ctmc} compares the modularisation using BDDs with the CTMC approach.
The new BDD approach solves nearly all DFTs within \SI{1}{\second} and outperforms the CTMC approach by several orders of magnitude.
Modularisation is therefore offering clear performance benefits compared to plain CTMC analysis.

Fig.~\ref{fig:dft_bdd_ctmcmod} compares the new BDD-based modularisation with the existing top-down modularisation.
The new approach prevails for all larger DFTs.
The BDD modularisation solved the adapted SFTs within \SI{0.1}{\second} while the top-down approach required up to \SI{100}{\second}.
On these DFTs, top-down modularisation is not applicable and thus, the entire DFT must be translated into a CTMC.

The advantage of the BDD modularisation becomes even clearer for multiple time bounds, \cf Fig.~\ref{fig:dft_bdd_vec_ctmcmod_vec}.
The BDD modularisation is significantly faster than the top-down modularisation for most of the considered DFTs.
On most DFTs, the BDD modularisation is able to compute \num{1000} time points within \SI{1}{\second}.
The main reasons for the performance improvement on multiple time points are the caching of the intermediate Markov models and the use of vectorisation on the resulting BDD, \cf Sect.~\ref{subsec:sft_unreliability}.

\noindent\emph{\textcolor{blue}{The BDD-based modularisation is significantly faster than both the plain CTMC approach and the existing top-down modularisation.}}

%% file: figures/dft_plots/dft_bdd_ctmc.tex
\begin{tikzpicture}
  \begin{axis}[scatterruntime,
    xmode=log,ymode=log,
    xlabel=\stormdft CTMC, ylabel=\stormdft BDD,
    legend pos=north west]
  \addplot[
    scatter,only marks,
    scatter/classes={
      mod-sft={modsft},
      mod-railway={modrailway},
      mod-vgs={modvgs},
      ffort={ffort}
    },
    scatter src=explicit symbolic]
    table [col sep=semicolon, x=CTMC_NOMOD, y=BDD, meta=Set]
    {figures/dft_plots/dft_plots.csv};
  \legend{Adapt. SFT, Adapt. Railway, Adapt. VGS, FFORT}
  \diagonalsTimes
  \end{axis}
\end{tikzpicture}

%% file: figures/dft_plots/dft_bdd_ctmcmod.tex
\begin{tikzpicture}
  \begin{axis}[scatterruntime,
    xmode=log,ymode=log,
    xlabel=\stormdft Top-down, ylabel=\stormdft BDD,
    legend pos=north west]
  \addplot[
    scatter,only marks,
    scatter/classes={
      mod-sft={modsft},
      mod-railway={modrailway},
      mod-vgs={modvgs},
      ffort={ffort}
    },
    scatter src=explicit symbolic]
    table [col sep=semicolon, x=CTMC_MOD, y=BDD, meta=Set]
    {figures/dft_plots/dft_plots.csv};
  \legend{Adapt. SFT, Adapt. Railway, Adapt. VGS, FFORT}
  \diagonalsTimes
  \end{axis}
\end{tikzpicture}

%% file: figures/dft_plots/dft_bdd_vec_ctmcmod_vec.tex
\begin{tikzpicture}
  \begin{axis}[scatterruntime,
    xmode=log,ymode=log,
    xlabel=\stormdft Top-down, ylabel=\stormdft BDD,
    legend pos=north west]
  \addplot[
    scatter,only marks,
    scatter/classes={
      mod-sft={modsft},
      mod-railway={modrailway},
      mod-vgs={modvgs},
      ffort={ffort}
    },
    scatter src=explicit symbolic]
    table [col sep=semicolon, x=CTMC_MOD_VEC, y=BDD_VEC, meta=Set]
    {figures/dft_plots/dft_plots.csv};
  \legend{Adapt. SFT, Adapt. Railway, Adapt. VGS, FFORT}
  \diagonalsTimes
  \end{axis}
\end{tikzpicture}

%% file: sections/conclusion.tex
\section{Conclusion}
\label{sec:conclusion}
We presented an implementation for fault tree analysis based on BDDs in the \stormdft tool.
Our implementation is competitive compared to existing tools and performs significantly better when computing multiple time points.
We also presented an implementation for DFT analysis based on modularisation.
The modular analysis allows to use the best techniques for each sub-tree and outperforms existing approaches.
\stormdft is currently the only available tool supporting modularisation for efficient DFT analysis.

\paragraph{Future work.}
Further improvements are needed to obtain smaller BDDs during the translation, for example by improving the heuristics for variable orderings such as using heuristics from \scram.
The modularisation cannot be fully exploited if large dynamic modules are present.
A possible research direction is to approximate the results for sub-modules, either by smaller fault trees or by the approximation approach of~\cite{DBLP:journals/tii/VolkJK18}.

%% file: sections/appendix.tex
\section{Additional details on evaluation of SFT approach}
\label{app:details_sft}

\subsection{Additional results}
\label{app:additional_results}

\paragraph{MCS with cut-off.}
\begin{figure}[t]
    \centering
    \subfigure[Runtime \stormdft vs \scram]{
        \centering
        \input{figures/sft_plots/storm_scram_mcs_cutoff}
        \label{fig:storm_scram_mcs_cutoff}
    }
    \subfigure[Runtime \stormdft vs \xfta]{
        \centering
        \input{figures/sft_plots/storm_xfta_mcs_cutoff}
        \label{fig:storm_xfta_mcs_cutoff}
    }
    \caption{Comparison for computation of MCS with cut-off}
    \label{fig:eval_mcs_cutoff}
\end{figure}
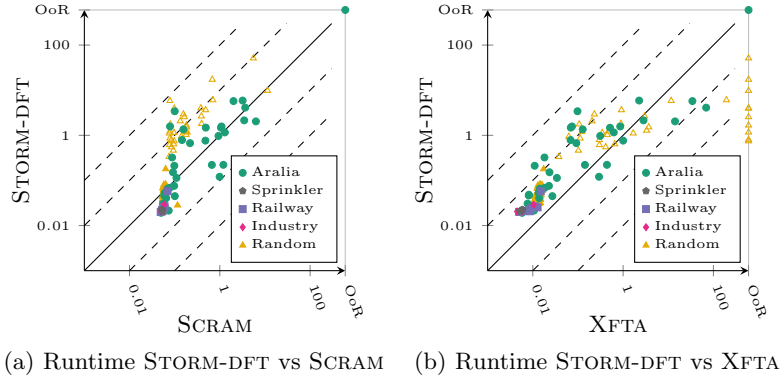
Fig.~\ref{fig:eval_mcs_cutoff} depicts the computation of the MCS with a given cut-off point to only print the most relevant cut sets.
In our setting, we restricted the computation to MCS with at most \num{4} \BE{}s.
As fewer cut sets need to be printed, the effect of the IO-operations is negligible.
The resulting performance is therefore very similar to the computation of the unreliability, \cf Fig.~\ref{fig:storm_scram_prob} and~\ref{fig:storm_xfta_prob}.

\paragraph{Multi-core computations.}
\begin{figure}[t]
    \centering
    \subfigure[Runtime for unreliability]{
        \input{figures/sft_plots/storm_mult_storm_prob}
        \label{fig:storm_mult_storm_prob}
    }
    \caption{Comparison single- vs multi-core in \stormdft}
    \label{fig:eval_multi_unrel_appendix}
\end{figure}
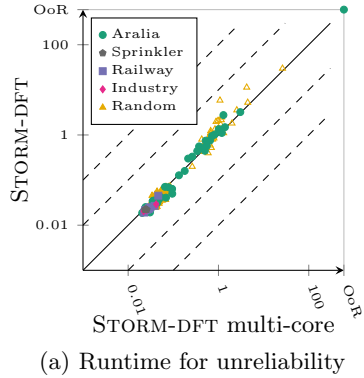
Fig.~\ref{fig:storm_mult_storm_prob} compares the effect of using multiple cores instead of a single core for computing the unreliability.
We draw the same conclusion as for Fig.~\ref{fig:storm_mult_storm_mcs}: using multiple cores has no visible performance advantage.

\paragraph{Importance measures.}
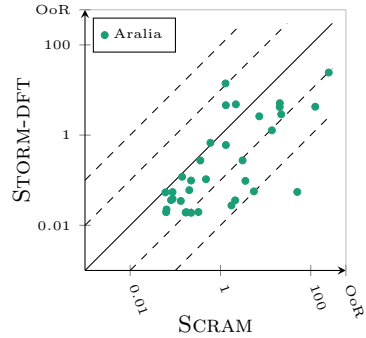
\begin{figure}[t]
    \centering
    \subfigure[Runtime \stormdft vs \scram]{
        \centering
        \input{figures/sft_plots/storm_scram_importance}
        \label{fig:storm_scram_importance}
    }
    \caption{Comparison for Birnbaum importance index}
    \label{fig:eval_scram_importance_appendix}
\end{figure}
Fig.~\ref{fig:storm_scram_importance} compares the runtime of \stormdft and \scram on the Aralia benchmarks when computing the Birnbaum importance index for all \BE{}s.
Note that \scram needs to compute the MCS in order to compute the importance index.
Thus, \stormdft is significantly faster than \scram.

\subsection{Reproducibility}
\paragraph{Artefact.}
We provide an artifact to reproduce our experiments\footnote{\url{https://doi.org/10.5281/zenodo.6390998}}.
The artifact contains the three tools, all SFT and DFT benchmark files, scripts to run the experiments as well as tables with detailed results.

\paragraph{Random SFTs.}
The 160 random SFT were generated using a script\footnote{\url{https://github.com/rakhimov/scram/blob/develop/scripts/fault_tree_generator.py}} provided by \scram.
We used the default settings of the script, but increased the number of BEs to obtain larger SFTs.
For the 128 random SFTs we set the number of \BE{}s to 150 and used seeds from $128$ to $255$ to initialize the random number generator for each SFT.
For the 33 large SFTs, we set the number of \BE{}s to 500 and used seeds from $127$ to $159$.

%% file: figures/sft_plots/storm_scram_mcs_cutoff.tex
\begin{tikzpicture}
  \begin{axis}[scatterruntime,
      xmode=log, ymode=log,
      xlabel=\scram, ylabel=\stormdft,
      legend pos=south east
    ]
  \addplot[
    scatter,only marks,
    scatter/classes={
      aralia={aralia},
      sprinkler={sprinkler},
      railway={railway},
      industry={industry},
      randomScram={randomScram},
      randomScramProb={randomScramProb}
    },
    scatter src=explicit symbolic]
    table [col sep=semicolon, x=SCRAM_MCS_CUT, y=STORM_MCS_CUT_DFS, meta=Set]
    {figures/sft_plots/sft_plots.csv};
  \legend{Aralia,Sprinkler,Railway,Industry,Random}
  \diagonalsTimes
  \end{axis}
\end{tikzpicture}

%% file: figures/sft_plots/storm_xfta_mcs_cutoff.tex
\begin{tikzpicture}
  \begin{axis}[scatterruntime,
      xmode=log, ymode=log,
      xlabel=\xfta, ylabel=\stormdft,
      legend pos=south east
    ]
  \addplot[
    scatter,only marks,
    scatter/classes={
      aralia={aralia},
      sprinkler={sprinkler},
      railway={railway},
      industry={industry},
      randomScram={randomScram},
      randomScramProb={randomScramProb}
    },
    scatter src=explicit symbolic]
    table [col sep=semicolon, x=XFTA_MCS_CUT, y=STORM_MCS_CUT_DFS, meta=Set]
    {figures/sft_plots/sft_plots.csv};
  \legend{Aralia,Sprinkler,Railway,Industry,Random}
  \diagonalsTimes
  \end{axis}
\end{tikzpicture}

%% file: figures/sft_plots/storm_mult_storm_prob.tex
\begin{tikzpicture}
  \begin{axis}[scatterruntime,
      xmode=log, ymode=log,
      xlabel=\stormdft multi-core, ylabel=\stormdft,
    ]
  \addplot[
    scatter,only marks,
    scatter/classes={
      aralia={aralia},
      sprinkler={sprinkler},
      railway={railway},
      industry={industry},
      randomScram={randomScram},
      randomScramProb={randomScramProb}
    },
    scatter src=explicit symbolic]
    table [col sep=semicolon, x=STORM_MULT_PROB_DFS, y=STORM_PROB_DFS, meta=Set]
    {figures/sft_plots/sft_plots.csv};
  \legend{Aralia,Sprinkler,Railway,Industry,Random}
  \diagonalsTimes
  \end{axis}
\end{tikzpicture}

%% file: figures/sft_plots/storm_scram_importance.tex
\begin{tikzpicture}
  \begin{axis}[scatterruntime,
      xmode=log, ymode=log,
      xlabel=\scram, ylabel=\stormdft,
    ]
  \addplot[
    scatter,only marks,
    scatter/classes={
      aralia={aralia}
    },
    scatter src=explicit symbolic]
    table [col sep=semicolon, x=SCRAM_IMP, y=STORM_IMP, meta=Set]
    {figures/sft_plots/scram_importance_plot.csv};
    \legend{Aralia}
  \diagonalsTimes
  \end{axis}
\end{tikzpicture}